\documentclass[aps,prb,twocolumn,amsmath,amssymb]{revtex4-1}
\usepackage{tabularx}
\usepackage{bm}
\usepackage{euscript}
\usepackage{graphicx}
\usepackage{color}
\usepackage{amsfonts}
\usepackage{exscale}
\usepackage{amsbsy}
\usepackage{subfigure}
\usepackage{textcomp}

\begin{document}

\title{Topological superconductivity and Majorana fermions in half-metal / superconductor heterostructure}

\author{Suk Bum Chung, Hai-Jun Zhang, Xiao-Liang Qi, and Shou-Cheng Zhang}
\affiliation{Department of Physics, Stanford University, Stanford, CA 94305}
\date{\today}

\begin{abstract}
As a half-metal is spin-polarized at its Fermi level by definition, it was conventionally thought to have little proximity effect to an $s$-wave superconductor. Here we show that 
with interface spin-orbit coupling 
$p_x +ip_y$ superconductivity without spin degeneracy is induced on the half-metal, and we give an estimate of its bulk energy gap. Therefore a single-band half-metal can give us a topological superconductor with a single chiral Majorana edge state. Our band calculation shows that two atomic layers of VTe or CrO$_2$ is a single-band half-metal for a wide range ($\sim$0.1eV) of Fermi energy and thus is a suitable candidate material.  
\end{abstract}

\maketitle

{\it Introduction:} Possibility of Majorana fermions arising out of condensed matter system has aroused great interest in recent years \cite{WILCZEK2009}. One class of systems where Majorana fermions can appear is the two-dimensional (2D) chiral superconductor which has a full pairing gap in the bulk, and $\mathcal{N}$ gapless chiral one-dimensional (1D) states, which consists of Majorana fermions \cite{Volovik1988, READ2000}, at the edge. In a $\mathcal{N}=1$ chiral topological superconductor (TSC), a single Majorana zero mode is bound to a vortex core \cite{JACKIW1981, Volovik1999, READ2000}, giving rise to non-Abelian statistics which can be potentially useful for topological quantum computation \cite{NAYAK2008}. The most straightforward way to realize such a chiral TSC is the intrinsic $p_x + ip_y$ superconductivity in spinless fermions \cite{READ2000}. The strongest candidate material for this superconductivity, albeit a spinful version, is Sr$_2$RuO$_4$ \cite{MACKENZIE2003}, but the experimental situation is not definitive \cite{Kam-edge}. Recently, there has been alternative proposals involving inducing $s$-wave superconductivity in material with strong spin-orbit coupling through proximity effect \cite{FU2008, SATO2009, lee2009, SAU2010, Qi2010a, lutchyn2010,alicea2010, OREG2010}. It was pointed out in one of the proposals \cite{lee2009} that spin polarized $p_x+ip_y$ superconductor can be obtained in a ferromagnetic film through proximity to a superconductor. In this Letter, we further develop this line of approach 
and demonstrate through explicit calculations the feasibility of creating Majorana fermions in a half metal / conventional $s$-wave superconductor heterostructure. 

We consider the pair formation on a half-metal (HM) that is in proximity contact to an $s$-wave superconductor (SC). A HM, by definition, is spin-polarized at the Fermi surface \cite{deGROOT1983}, {\it i.e.} it is a metal for the majority-spin and an insulator for the minority-spin. Our proposal has two major advantages over other current proposals. Firstly, the $\mathcal{N}=1$ TSC phase in our approach exists in a wide range of Fermi level. Secondly, we expect the proximity effect between SC and HM to be more robust than that between SC and semiconductor, due to better Fermi surface matching.
It has been known that at the normal metal to $s$-wave SC interface, $p$-wave pairing can be induced because inversion symmetry is broken \cite{Gorkov2001}. Eschrig {\it et al.} showed that when normal metal is HM, even frequency pairing would be mostly $p$-wave \cite{ESCHRIG2003}. Furthermore, there are experimental indications of 
strong proximity effect between a HM and an $s$-wave superconductor 
\cite{Keizer2006, Kalcheim2010}. Here we will show how we can obtain $p_x+ip_y$ pairing symmetry in a 2D HM when it is coupled to an $s$-wave superconductor only through electron hopping across the interface. 
If the 2D HM has a single Fermi pocket without spin degeneracy, such $p_x+ip_y$ pairing will give us the TSC with $\mathcal{N} = 1$. We will show band calculation for thin film material that is HM and has a single Fermi pocket. We will also discuss the suitable superconductor for optimizing this proximity effect and the method we can use for detecting the $p_x+ip_y$ pairing in the HM.

\begin{figure}
\centerline{\includegraphics[width=.20\textwidth]{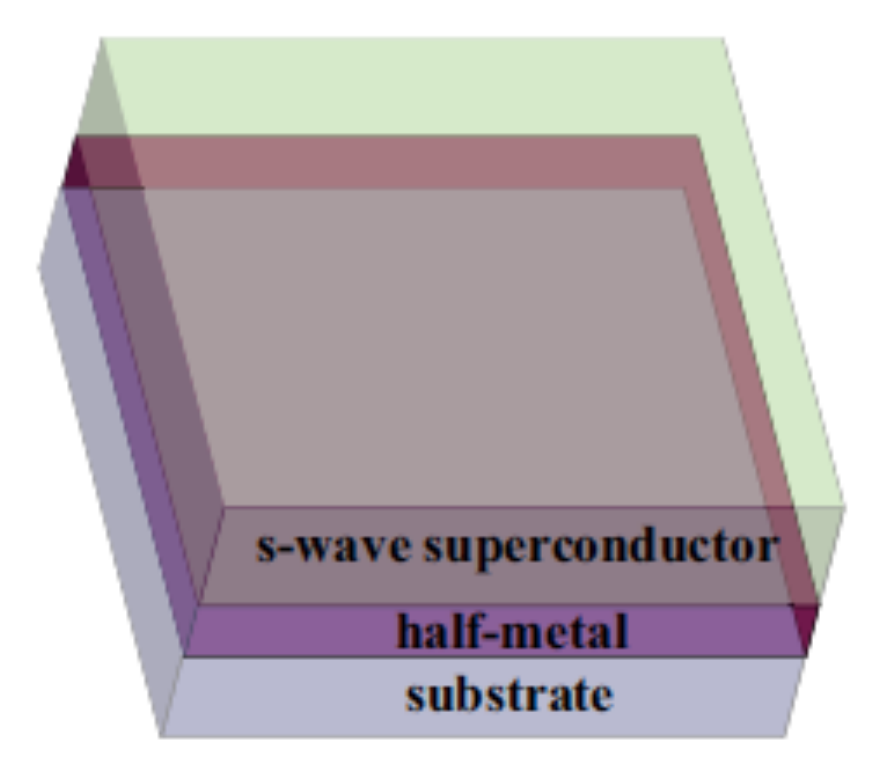}}
\caption{The heterostructure for obtaining $p_x + ip_y$ superconductivity. The half-metal layer has a 2D electronic structure with a single Fermi surface without spin degeneracy. It is coupled to the $s$-wave superconductor through hopping. The substrate stabilizes the half-metal crystal structure without affecting qualitatively its electronic structure.}
\label{FIG:hopping}
\vspace{-5mm}
\end{figure}

{\it Basic Model:} 
We consider the model with a $s$-wave superconductor and a 2D half-metal coupled by a weak hopping between two systems:
\begin{equation}
\mathcal{H} = \mathcal{H}_{SC} + \mathcal{H}_{HM} + \mathcal{H}_t
\label{EQ:model}
\end{equation}
where
\begin{align}
\mathcal{H}_{SC} =& \sum_{{\bf k},\sigma} (\epsilon'_{\bf k} - \mu') c^\dagger_{{\bf k}\sigma} c_{{\bf k}\sigma} + \sum_{\bf k} (\Delta'_{\bf k} c^\dagger_{{\bf k}\uparrow} c^\dagger_{-{\bf k}\downarrow}  + {\rm h.c.}),\nonumber\\
\mathcal{H}_{HM} =& \sum_{\bf k} (\epsilon_{\bf k} - \mu) f^\dagger_{{\bf k}\uparrow} f_{{\bf k}\uparrow},\nonumber\\
\mathcal{H}_t =& \sum_{{\bf k}\sigma} (t_{{\bf k},\uparrow\sigma} f^\dagger_{{\bf k}\uparrow} c_{{\bf k}\sigma} + {\rm h.c.})
\label{EQ:hopModel}
\end{align}
(note ${\bf k}$ is an in-plane vector). 
This model provides a general mechanism for SC proximity effect. Note that while $\mathcal{H}_t$ requires ${\bf k}$ to be conserved in the hopping process, it allows for both spin-flip hopping and dependence of hopping on ${\bf k}$. 

In this model, the symmetry of Cooper pairs formed on the half-metal side, $\langle f_{-{\bf k}\uparrow} f_{{\bf k}\uparrow}\rangle$, is determined entirely by $t_{{\bf k},\uparrow\sigma}$. In this model, the only channel for this Cooper pair formation is to have the HM electrons with momenta ${\bf k}$ and $-{\bf k}$ hop to the SC side to form a pair there. This process requires that hopping flips the spin of either one of (but not both) the ${\bf k}$ and $-{\bf k}$ electrons. Since the Cooper pair on the SC side is in the spin-singlet $s$-wave state, the two hopping amplitudes interferes destructively, giving us
\begin{equation}
\langle f_{-{\bf k}\uparrow} f_{{\bf k}\uparrow} \rangle \propto t_{{\bf k},\uparrow\uparrow} t_{-{\bf k},\uparrow\downarrow} - t_{-{\bf k},\uparrow\uparrow} t_{{\bf k},\uparrow\downarrow}
\label{EQ:hoppingPair}
\end{equation}
with an $s$-wave multiplicative factor; 
note the odd spatial parity. 
In the limit of weak hopping, $|t_{{\bf k},\uparrow\sigma}| \ll |\Delta'_{\bf k}|$, we find the pairing amplitude at the HM Fermi surface ({\it i.e.} for ${\bf k}$ that satisfies $\epsilon_{\bf k} - \mu = 0$) to be \cite{Suppl}
\begin{equation}
\langle f_{-{\bf k}\uparrow} f_{{\bf k}\uparrow} \rangle \approx \frac{1}{2}\frac{\eta_{\bf k}}{\sqrt{|\eta_{\bf k}|^2+|\zeta_{\bf k}|^2}},
\label{EQ:PairFS}
\end{equation}
where
\begin{align}
\eta_{\bf k} =& (t_{{\bf k},\uparrow\uparrow} t_{-{\bf k},\uparrow\downarrow} - t_{{\bf k},\uparrow\downarrow} t_{-{\bf k},\uparrow\uparrow})\langle c_{-{\bf k}\downarrow} c_{{\bf k}\uparrow} \rangle|_{\epsilon_{\bf k} = \mu},\nonumber\\
\zeta_{\bf k} = & \frac{|t_{{\bf k},\uparrow\uparrow}|^2 + |t_{{\bf k},\uparrow\downarrow}|^2}{2}\frac{\epsilon'_{\bf k} - \mu'}{E'_{\bf k}}\vert_{\epsilon_{\bf k} = \mu}
\label{EQ:PairFS2}
\end{align}
with $E'_{\bf k} = \sqrt{(\epsilon'_{\bf k} - \mu')^2 + |\Delta'_{\bf k}|^2}$ ($|t_{{\bf k},\uparrow\sigma}|^2 = |t_{-{\bf k},\uparrow\sigma}|^2$ assumed).
Eqs. \eqref{EQ:PairFS} and \eqref{EQ:PairFS2} tell us that if the Fermi surfaces of the HM and the SC match exactly ({\it i.e.} $\epsilon'_{\bf k} - \mu' = 0$ when $\epsilon_{\bf k} - \mu = 0$), we have $\langle f_{-{\bf k}\uparrow} f_{{\bf k}\uparrow} \rangle = e^{i\phi_{\bf k}}\langle c_{-{\bf k}\downarrow} c_{{\bf k}\uparrow} \rangle$ at the HM Fermi surface, with $e^{i\phi_{\bf k}}$ being the phase factor of $t_{{\bf k},\uparrow\uparrow} t_{-{\bf k},\uparrow\downarrow} - t_{{\bf k},\uparrow\downarrow} t_{-{\bf k},\uparrow\uparrow}$, . Physically, 
$\eta_{\bf k}$ is proportional to the amplitude that the ${\bf k}$ and $-{\bf k}$ HM electrons hop to the $s$-wave SC with opposite spins and form a Cooper pair, while $\zeta_{\bf k}$ is proportional to the amplitude that these electrons hop to the SC with same spins aligned and therefore do not form a Cooper pair.

\begin{figure}
\centerline{\includegraphics[width=.39\textwidth]{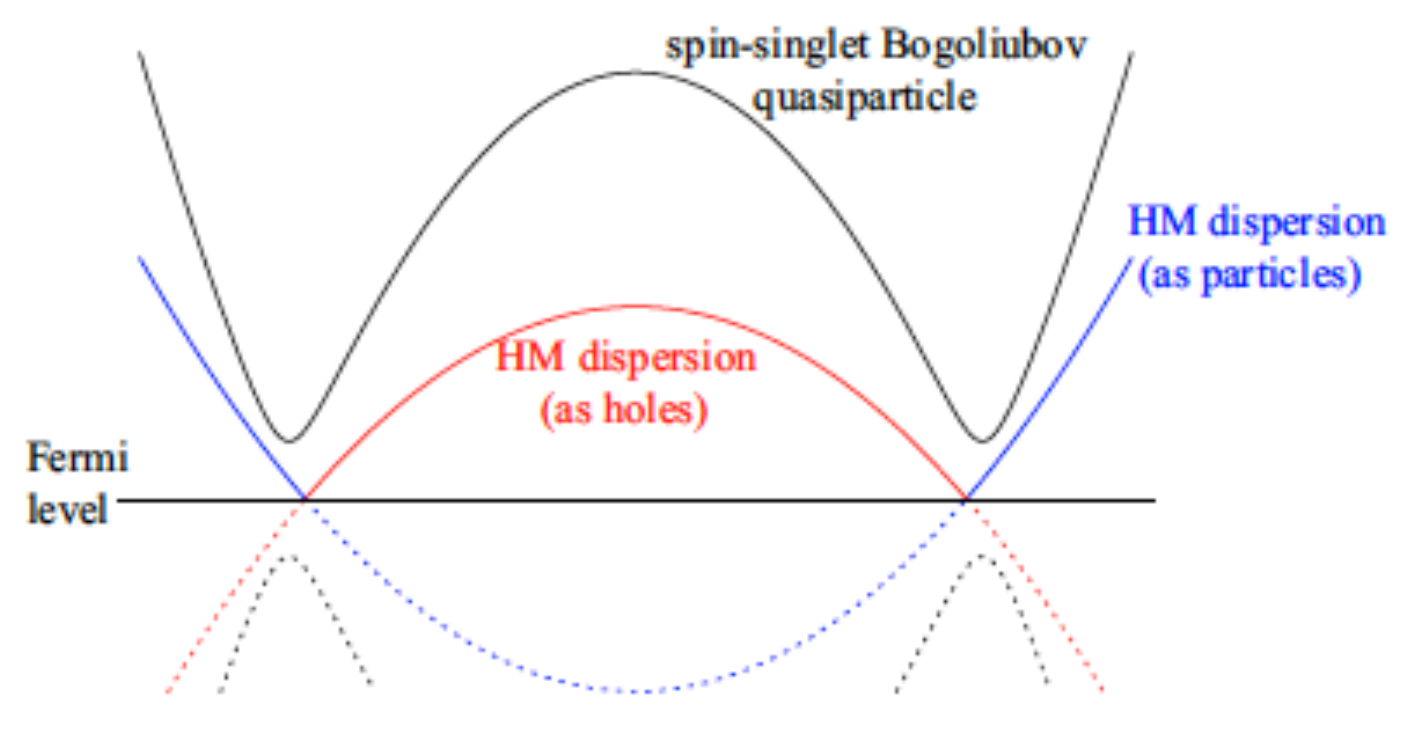}}
\caption{Schematic representation of the HM and SC band for weak hopping. Two HM bands, with negative energy portions dotted, are due to the artificial doubling of degrees of freedom in the Bogoliubov-de Gennes formalism. In this formalism, the pairing amplitude and gap require hybridization of the `particle' and `hole' bands, which can only occur through hopping to the SC.}
\label{FIG:hopBand}
\vspace{-5mm}
\end{figure}

Our model gives us not only the pairing amplitude but also the pairing gap on the HM side. In the weak hopping limit we have been discussing, we obtain
\begin{equation}
\Delta^{HM}_{\bf k} = \frac{\eta_{\bf k}}{E'_{\bf k}} = \frac{t_{{\bf k},\uparrow\uparrow} t_{-{\bf k},\uparrow\downarrow} - t_{{\bf k},\uparrow\downarrow} t_{-{\bf k},\uparrow\uparrow}}{2E'^2_{\bf k}}\Delta'_{\bf k}
\label{EQ:HMgap0}
\end{equation}
at the HM Fermi surface \cite{Suppl}. Note here that $|\Delta'_{\bf k}|$ is maximized when we have a perfect Fermi surface matching between the HM and the SC. We point out that since our HM is 2D, proximity to a SC leads to a pairing gap \cite{DEUTSCHER2005} \cite{footnote}.
Eqs.\eqref{EQ:PairFS} and \eqref{EQ:HMgap} tells us this gap $\Delta^{HM}$ comes from the HM electron pairing. We therefore conclude that the topological property of SC induced in the HM to be determined entirely by $\Delta^{HM}$, or equivalently, by $t_{{\bf k},\uparrow\sigma}$. 

In the limit of strong hopping, we obtain a much larger pairing gap while the pairing symmetry is still determined by the hopping as in Eq.\eqref{EQ:hoppingPair}. Our strong hopping limit requires at the HM Fermi surface $\epsilon_{\bf k} - \mu = 0$, the energetics is dominated by the spin-conserving hopping $t_{{\bf k},\uparrow\uparrow}$, {\it i.e.} $|t_{{\bf k},\uparrow\uparrow}| \gg E'_{\bf k}$ and $|t_{{\bf k},\uparrow\uparrow}| \gg |t_{{\bf k},\uparrow\downarrow}|$. Within this model, at the HM Fermi surface ($\epsilon_{\bf k} - \mu =0$), while the pairing amplitude  remains that of Eq.\eqref{EQ:PairFS}, the pairing gap at the HM is now given as \cite{Suppl}
\begin{equation}
\Delta^{HM}_{\bf k} = \frac{\eta_{\bf k}E'_{\bf k}}{|t_{{\bf k},\uparrow\uparrow}|^2} = \frac{t_{{\bf k},\uparrow\uparrow} t_{-{\bf k},\uparrow\downarrow} - t_{{\bf k},\uparrow\downarrow} t_{-{\bf k},\uparrow\uparrow}}{2|t_{{\bf k},\uparrow\uparrow}|^2} \Delta'_{\bf k}.
\label{EQ:HMgap}
\end{equation}
The HM pairing gap in the strong hopping limit, given in Eq.\eqref{EQ:HMgap}, is much larger than that in the weak hopping limit, given in Eq.\eqref{EQ:HMgap0}. 
This is because in Eq.\eqref{EQ:HMgap}, $\Delta^{HM}$ is proportional to $t_{\uparrow\downarrow}/t_{\uparrow\uparrow}$, while in Eq.\eqref{EQ:HMgap0}, it is proportional to $t_{\uparrow\downarrow} t_{\uparrow\uparrow}/E'^2$. 
Similar analysis can be applied to the proximity of HM with a 3D SC, in which case $\Delta_{\bf k}^{HM}$ is still linearly proportional to $t_{\uparrow\downarrow}$.

{\it Interface hopping:} To show how this HM / SC proximity effect gives us the $p_x + ip_y$ pairing, we first discuss the symmetry $t_{{\bf k},\uparrow\sigma}$. We note that, since inversion symmetry is broken at the HM/SC interface, there will be an interface Rashba spin-orbit coupling (SOC) \cite{Gorkov2001, lee2009}:
\begin{equation}
\mathcal{H}_{SOC} = \hbar \alpha ({\bm \sigma} \times {\bf k})\cdot {\bf \hat{n}}\delta({\bf \hat{n}} \cdot {\bf r}).
\label{EQ:interSOC}
\end{equation}
with $\hat{\bf n}$ the normal direction of the interface. To account for this interface SOC in our model Eq.\eqref{EQ:hopModel}, we need to have a hopping term that has the same symmetry as Eq.\eqref{EQ:interSOC}. Such a hopping term, in the momentum space, can be written as
\begin{equation}
\mathcal{H}_{t-SOC} =   t_{SOC} \sum_{\bf k} F^\dagger_{\bf k} (\sigma^x \sin k_y a - \sigma^y \sin k_x a) C_{\bf k} +{\rm h.c.},
\label{EQ:hopSOC}
\end{equation}
where $F_{\bf k} = (f_{{\bf k}\uparrow}, f_{{\bf k}\downarrow})^T$ and $C_{\bf k} = (c_{{\bf k}\uparrow}, c_{{\bf k}\downarrow})^T$. However, the terms involving $f_{{\bf k}\downarrow}$ can be ignored for the HM/SC interface, as these terms involve process occurring at energy larger than the minority-spin gap of the HM.

We can now show explicitly how we obtain the $p_x + ip_y$ pairing from $t_{{\bf k},\uparrow\sigma}$. 
Assuming the spin-conserving hopping to be momentum-independent, we obtain
\begin{align}
t_{{\bf k},\uparrow\uparrow} =& t_0\nonumber\\
t_{{\bf k},\uparrow\downarrow} =& t_{SOC} (i\sin k_x a + \sin k_y a).
\label{EQ:hopSymm}
\end{align}
We note that for a square lattice in the real space representation, the spin-conserving hopping of Eq.\eqref{EQ:hopSymm} is `vertical', {\it i.e.} $t_0\sum_i f^\dagger_{i\uparrow} c_{i\uparrow}$ +  h.c., while the spin-flip hopping is of the `nearest neighbor' type, {\it i.e.} $t_{SO} \sum_{\langle ij \rangle} \exp[i\theta_{ij}] f^\dagger_{i\uparrow} c_{j\downarrow}$ + h.c., where $\theta_{ij}$ gives the chiral $p$-wave symmetry to this spin-flip hopping. Inserting these $t_{{\bf k},\uparrow\sigma}$ into Eq.\eqref{EQ:hoppingPair} gives us the chiral $p$-wave pairing on the HM side:
\begin{eqnarray}
\langle f_{-{\bf k}\uparrow} f_{{\bf k}\uparrow} \rangle &\propto& t_{{\bf k},\uparrow\uparrow} t_{-{\bf k},\uparrow\downarrow} - t_{-{\bf k},\uparrow\uparrow} t_{{\bf k},\uparrow\downarrow}\nonumber\\
&=& -2i t_0 t_{SOC}(\sin k_x a - i \sin k_y a).
\label{EQ:hopSymm2}
\end{eqnarray}
Eq.\eqref{EQ:hopSymm} gives us the $\mathcal{N} = 1$ TSC in the strong hopping limit as well as the weak hopping limit \cite{Suppl}. We also note that the chiral $p$-wave is mentioned as the likely pairing symmetry if there is intrinsic SC in a HM \cite{PICKETT1996}.

From the origin of the interface SOC, we can estimate of the HM pairing gap to be $|\Delta^{HM}| \sim |\Delta'|(\alpha_{SOC}/W)$, where $\alpha_{SOC}$ is the Rashba SOC of the $s$-wave SC and $W$ is the bandwidth. Physically, when SOC is strong for the $s$-wave SC but weak for the HM, we can expect to have the interface SOC. This situation is experimentally relevant because $s$-wave SC can exist in materials with strong SOC, while the spin-polarized ARPES indicates complete spin polarization for the HM, as in CrO$_2$ \cite{DEDKOV2002}. Assuming that we have zero effective SOC on the HM, we will effectively have the spin-orbit coupling hopping $t_{SOC} \sim (\alpha_{SOC}/W) t_0$ induced through second order perturbation. This is sufficient for estimating $|\Delta^{HM}|$, because, in the strong hopping limit, we find $\Delta^{HM} \sim (t_{SOC}/t_0)\Delta'$ by inserting Eq.\eqref{EQ:hopSymm2} into Eq.\eqref{EQ:HMgap}.

We also point out that this HM/SC proximity effect can provide us with a means to obtain a multi-domain chiral $p$-wave SC. Eqs.\eqref{EQ:hopSOC} and \eqref{EQ:hopSymm} show us that if we reverse the HM spin polarization, than we will also reverse the chirality of the induced SC. Therefore in a HM, a domain boundary between opposite spin polarization will also be the domain boundary between the $p_x+ip_y$ and $p_x-ip_y$ domain when SC is induced.

\begin{figure}
\centerline{\includegraphics[width=0.48\textwidth]{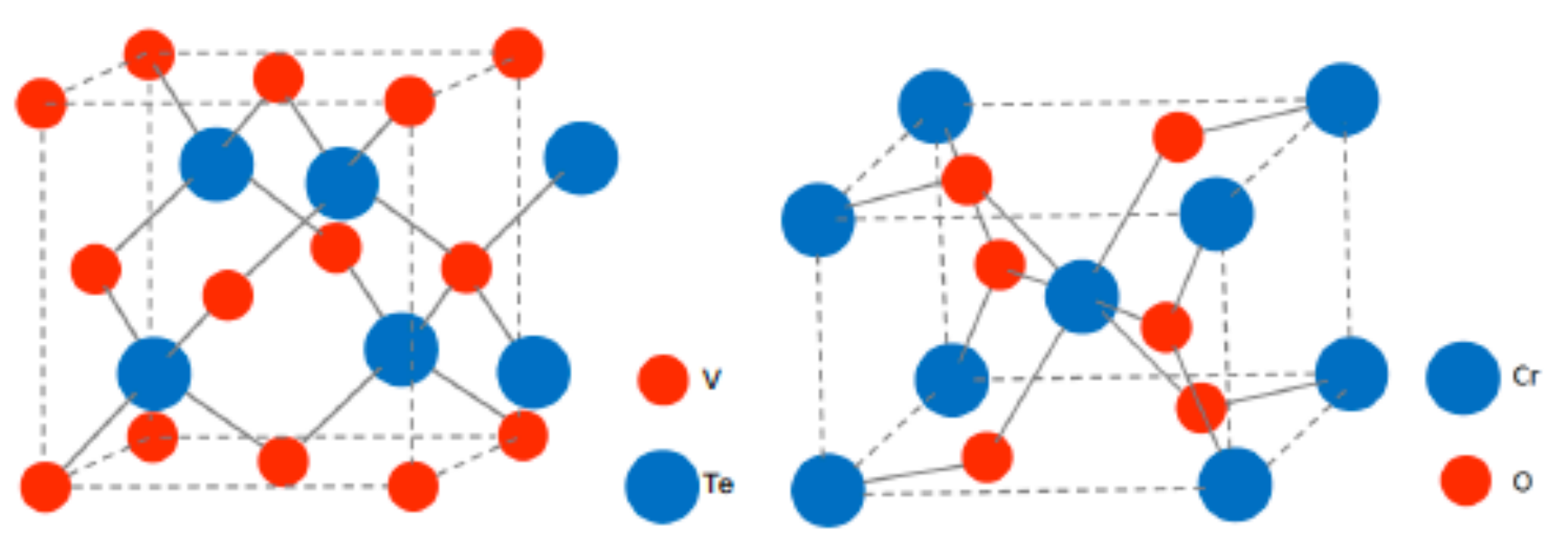}}
\caption{Crystal structures of two candidate materials, VTe and CrO$_2$.The left shows the zinc blende crystal structure of bulk VTe. Here, both V and Te forms body-centered cubic with the lattice constant of $a = 0.6271$nm. (The zinc blende CrTe has the same structure as the zinc blend VTe, but with a slightly smaller lattice constant 
.)  The left shows the rutile crystal structure of bulk CrO$_2$. The lattice constants are $a = 0.4421$nm and $c = 0.2916$nm. The distance between the nearest O and Cr on the same layer is 0.1817nm. 
}
\label{FIG:BulkCrO2}
\vspace{-5mm}
\end{figure}

{\it Candidate material:} We note that the spin-polarized (or single-spin) $p_x + ip_y$ SC with a single Fermi pocket gives us the $\mathcal{N} = 1$ TSC \cite{READ2000}, but the same cannot be said for the $p_x + ip_y$ SC with multiple Fermi pockets \cite{Raghu2010, QI2010}. Therefore to obtain the $\mathcal{N}=1$ TSC, it is important that we find a 2D HM with a single Fermi pocket, as we have in Eq.\eqref{EQ:hopModel}. That way, the HM / SC proximity effect we discussed will give us an equivalent of the single-spin $p_x + ip_y$ SC with a single Fermi pocket

\begin{figure}
\centerline{\includegraphics[width=0.48\textwidth]{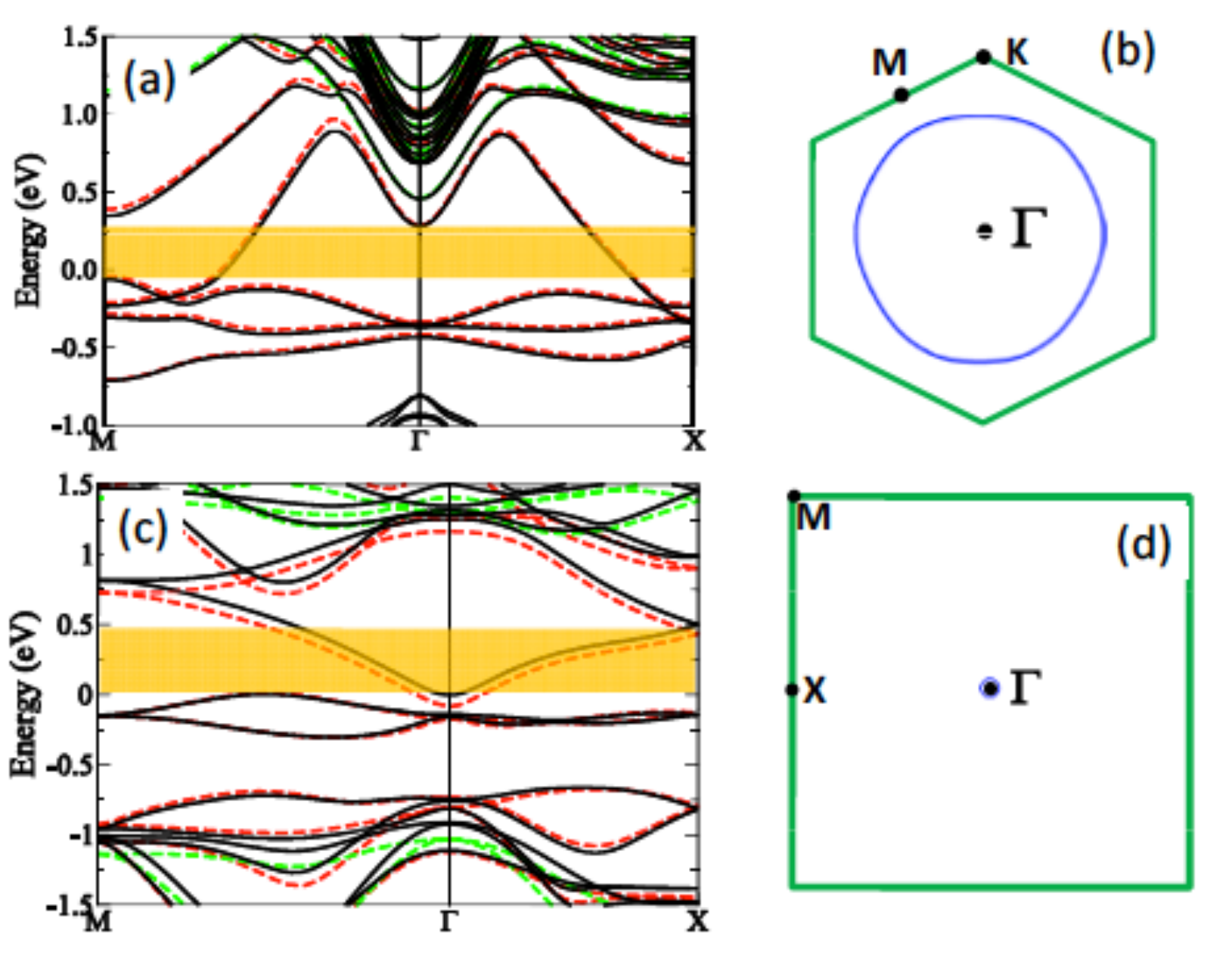}}
\caption{(a) The band structure of the VTe-ZnTe(111) model. There is no spin degeneracy in these bands; the red and the green dotted curve shows the spin-up and -down bands when SOC is absent. The region shaded in orange shows the energy range for which we obtain a single Fermi pocket. 
(b) The single Fermi surface at the energy level 0.0eV for ZnTe-VTe(111) model. The green box shows the first Brillouin zone(BZ). (c) The band structure of the CrO$_2$ (001) model, done in the same way as the VTe-ZnTe(111) model.  (d) The single Fermi surface around the $\Gamma$ point at the energy level 0.01eV for the CrO$_2$ (001) model (the green box showing the 1st BZ). } 
\label{FIG:bandFS}
\vspace{-5mm}
\end{figure}

One candidate material for a 2D HM with a single Fermi surface is the zinc blende VTe or CrTe that is two atomic layers thick in the (111) direction. 
We 
identified candidate material through {\it ab-initio} band calculations 
performed 
in the frame of density-functional theory\cite{hohenberg1964,kohn1965} with the plane-wave pseudopotential
method \cite{Suppl}. 
Both VTe and CrTe in the zinc blende structure have been shown to be half-metallic in band calculation \cite{XIE2003}; a thin film of the zinc blende CrTe has been fabricated in thin films by molecular-beam epitaxy \cite{SREENIVASAN2008}. As we see in Fig.~\ref{FIG:bandFS}, two atomic layers of zinc blende VTe (111) on the zinc blende ZnTe substrate 
is a half-metal with a single Fermi pocket at the Fermi level; this is due to 
a mechanism 
\cite{Suppl} analogous to the ``polar catastrophe" in the LaAlO$_3$ / SrTiO$_3$ interface \cite{HARRISON1978, POPOVIC2008}. 
As we have a 0.3eV range for the Fermi level that gives us a single Fermi pocket, unlike in many of the previous proposals for obtaining the $\mathcal{N}=1$ TSC \cite{SAU2010} we do not require fine-tuning of the HM Fermi level. For CrTe, we find a narrower Fermi level range for a single Fermi surface ($\approx$ 0.06eV).

We also point out that the CrO$_2$ film that is two atomic layers thick in the (001) direction comes close to fulfilling our requirement. Bulk CrO$_2$ has been experimentally confirmed to be half-metallic 
\cite{DEDKOV2002,SOULEN1998}, and it was also shown in experiment to have a strong proximity effect to an $s$-wave SC - NbTiN - that has a relatively high $T_c$ ($\sim 14$K) and a spin-orbit coupling larger than the SC gap. Since the width of bands near Fermi level for NbN is about 7.2eV \cite{MATTHEISS1972} and the atomic spin-orbit coupling of Nb is 0.1eV \cite{MACHIN1969}, the mechanism discussed here may be able to induce a pairing gap up to $\sim$1K.
The bulk CrO$_2$ crystal, as shown in Fig.~\ref{FIG:BulkCrO2}, has a rutile structure, where the Cr atoms form body-centered tetragonal lattice and are surrounded by distorted O octahedra \cite{KOROTIN1998}.  The lattice is compressed in the $c$-axis direction ($c/a = 0.66$), with the lattice constant $a = 0.44$nm. 
The band calculation for two atomic layer of CrO$_2$ (001) in a perfect rutile structure gives us a half-metal with one particle pocket around the $\Gamma$ point and four hole pockets, each with area 1/4 that of the particle pockets; as shown in Fig.~\ref{FIG:bandFS}, we can make hole pockets vanish by raising the energy level by less than 0.01eV. Note that the bandwidth if very narrow for hole pockets and we obtain a HM with a single Fermi surface for a Fermi level range of $\sim 0.5$eV. 

{\it Detection:} Detecting the chiral Majorana edge state along with the fully gapped quasiparticle spectrum in the bulk of the HM will confirm that we have $\mathcal{N} = 1$ TSC in the HM. The most direct method for detection would be measuring with STM the tunneling $dI/dV$ vs. $V$ spectra of the HM; indeed this has been used recently to detect the SC proximity effect in a HM \cite{Kalcheim2010}. If $p_x+ip_y$ pairing is induced, $dI/dV$ is suppressed below $V=|\Delta^{HM}|/e$ in the bulk while along the edge we will see the zero bias conductance peak. Another place to look for the zero bias conductance peak is the HM domain boundary. This is because this will also be the boundary between $p_x+ip_y$ and $p_x-ip_y$ domain in the SC state and there will be two chiral Majorana states running along it.

In summary, we have shown that we can obtain the $\mathcal{N} = 1$ TSC in a HM through proximity effect with an $s$-wave SC. In a model where the HM is coupled to the SC through hopping, the symmetry of SC pairing induced in the HM is determined entirely by the hopping term. Due to the interface Rashba spin-orbit coupling, the hopping term will induce the pairing with $p+ip$ symmetry. In order for this $p_x+ip_y$ pairing to lead to the $\mathcal{N} = 1$ TSC, we need to have a HM with single Fermi surface, and our band calculation shows that this can be obtained for a very thin CrO$_2$ film. STM measurement can be used to verify that TSC is induced in the HM. 

We would like to thank Mac Beasley, Richard Martin and Patrick Lee for sharing their insights; we also acknowledge helpful discussions with Steve Kivelson, Srinivas Raghu, and Kookrin Char. This work is supported by DOE under contract DE-AC02-76SF00515 (SBC), the Sloan Foundation (XLQ) and NSF under grant number DMR-0904264.

\appendix

\section{HM pairing correlation}

Our model Hamiltonian can be solved exactly, as it is quadratic:
\begin{align}
\mathcal{H} =& {\rm const.} + \sum_{\bf k} (\epsilon_{\bf k} - \mu) f^\dagger_{{\bf k}\uparrow} f_{{\bf k}\uparrow} + \sum_{{\bf k}\sigma}E_{\bf k}\gamma^\dagger_{{\bf k}\sigma} \gamma_{{\bf k}\sigma} \nonumber\\
&+ \sum_{\bf k}[f^\dagger_{{\bf k}\uparrow}\{t_{{\bf k},\uparrow\uparrow}(u^*_{\bf k}\gamma_{{\bf k}\uparrow} + v_{\bf k} \gamma^\dagger_{-{\bf k}\downarrow})\nonumber\\
&+ t_{{\bf k},\uparrow\downarrow}(u^*_{\bf k}\gamma_{{\bf k}\downarrow} - v_{\bf k} \gamma^\dagger_{-{\bf k}\uparrow})\}+{\rm h.c.}]\nonumber\\
=& \frac{1}{2}\sum_{\bf k} \Psi^\dagger_{\bf k} H^{BdG}_{\bf k} \Psi_{\bf k},
\end{align}
where $\Psi = (f_{{\bf k}\uparrow}, f^\dagger_{-{\bf k}\uparrow}, \gamma_{{\bf k}\uparrow}, \gamma_{{\bf k}\downarrow}, \gamma^\dagger_{-{\bf k}\uparrow}, \gamma^\dagger_{-{\bf k}\downarrow})^T$ and
\begin{equation}
H^{BdG}_{\bf k} = \left[\begin{array}{cc}
h^{HM} & T_{\bf k}\\
T^\dagger_{\bf k} & \Lambda_{BdG}
\end{array}\right],
\end{equation}
with $h^{HM} = \tau^z (\epsilon_{\bf k} - \mu)$, $\Lambda_{BdG} = {\rm diag}(E_{\bf k}, E_{\bf k}, -E_{\bf k}, -E_{\bf k})$, and
\begin{equation}
T_{\bf k} = \left[\begin{array}{cccc}
u^*_{\bf k} t_{{\bf k},\uparrow\uparrow} & u^*_{\bf k} t_{{\bf k},\uparrow\downarrow} &
-v_{\bf k} t_{{\bf k},\uparrow\downarrow} & v_{\bf k} t_{{\bf k},\uparrow\uparrow}\\
v^*_{\bf k} t^*_{-{\bf k},\uparrow\downarrow} & -v^*_{\bf k} t^*_{-{\bf k},\uparrow\uparrow}  &
-u_{\bf k} t^*_{-{\bf k},\uparrow\uparrow} & -u_{\bf k} t^*_{-{\bf k},\uparrow\downarrow}
\end{array}\right]
\end{equation}
This Hermitian matrix can be diagonalized: $U_{\bf k}^\dagger H^{BdG}_{\bf k} U_{\bf k} = \Lambda_{\bf k}$, where $\Lambda_{\bf k} = {\rm diag}(\lambda_1, \ldots \lambda_6)$ and $U_{\bf k}$ is a unitary matrix that becomes an identity matrix in the $t_{{\bf k},\uparrow\sigma} \to 0$ limit. We can also defined the basis $\tilde{\Psi}_{\bf k} \equiv U_{\bf k}^\dagger \Psi_{\bf k} = (\tilde{f}_{{\bf k}\uparrow}, \tilde{f}^\dagger_{-{\bf k}\uparrow}, \tilde{\gamma}_{{\bf k}\uparrow}, \tilde{\gamma}_{{\bf k}\downarrow}, \tilde{\gamma}^\dagger_{-{\bf k}\uparrow}, \tilde{\gamma}^\dagger_{-{\bf k}\downarrow})^T$ that diagonalize our BdG Hamiltonian.

We can obtain pairing correlator $\langle f_{-{\bf k}\uparrow} f_{{\bf k}\uparrow} \rangle$ from this diagonalization. Since this diagonalization allows us to express the original operators in terms of operators of $\tilde{\Psi}$,
\begin{align}
f_{{\bf k}\uparrow} =& (U_{\bf k})_{11} \tilde{f}_{{\bf k}\uparrow} + (U_{\bf k})_{12} \tilde{f}^\dagger_{-{\bf k}\uparrow} + (U_{\bf k})_{13} \tilde{\gamma}_{{\bf k}\uparrow}\nonumber\\
&+ (U_{\bf k})_{14} \tilde{\gamma}_{{\bf k}\downarrow}+ (U_{\bf k})_{15}  \tilde{\gamma}^\dagger_{-{\bf k}\uparrow} + (U_{\bf k})_{16}  \tilde{\gamma}^\dagger_{-{\bf k}\downarrow}\nonumber\\
f^\dagger_{-{\bf k}\uparrow} =& (U_{\bf k})_{21} \tilde{f}_{{\bf k}\uparrow} + (U_{\bf k})_{22} \tilde{f}^\dagger_{-{\bf k}\uparrow} + (U_{\bf k})_{23} \tilde{\gamma}_{{\bf k}\uparrow}\nonumber\\
&+ (U_{\bf k})_{24} \tilde{\gamma}_{{\bf k}\downarrow}+ (U_{\bf k})_{25}  \tilde{\gamma}^\dagger_{-{\bf k}\uparrow} + (U_{\bf k})_{26}  \tilde{\gamma}^\dagger_{-{\bf k}\downarrow},
\end{align}
we can express the pairing amplitude in terms of elements of $U_{\bf k}$:
\begin{align}
\langle f_{-{\bf k}\uparrow} f_{{\bf k}\uparrow} \rangle =& (U_{\bf k})_{12} (U_{\bf k})^*_{22} + (U_{\bf k})_{15} (U_{\bf k})^*_{25} + (U_{\bf k})_{16} (U_{\bf k})^*_{26}\nonumber\\
=& (U_{\bf k})^*_{21} (U_{\bf k})_{22} + (U_{\bf k})^*_{51} (U_{\bf k})_{52} + (U_{\bf k})^*_{61} (U_{\bf k})_{62}.
\end{align}
Since the column vectors of $U_{\bf k}$ are eigenvectors of $H^{BdG}_{\bf k}$, we can obtain the pairing amplitude by obtaining the eigenvectors of $H^{BdG}_{\bf k}$. In practice, both the weak and strong hopping limit enables us to use perturbation theory for calculating the pairing amplitude.

\subsection{Weak hopping at the HM Fermi surface}

Here, we assume $|t_{{\bf k},\uparrow\sigma}| \ll |\Delta_{\bf k}|$.
At the HM Fermi surface, we need to apply the degenerate second order perturbation theory since $h^{HM} = 0$. 
Therefore, we write down eigenvectors only up to terms linear in $t_{{\bf k}\sigma}$:
\begin{equation}
{\bf F}^{(i)}_{\bf k} = \left[\begin{array}{c} A^{(i)} \\ B^{(i)} \\ -[A^{(i)} u_{\bf k}t^*_{{\bf k},\uparrow\uparrow} + B^{(i)} v_{\bf k}t_{-{\bf k},\uparrow\downarrow}]/E_{\bf k} \\ -[A^{(i)} u_{\bf k}t^*_{{\bf k},\uparrow\downarrow} - B^{(i)} v_{\bf k}t_{-{\bf k},\uparrow\uparrow}]/E_{\bf k} \\ -[A^{(i)} v^*_{\bf k}t^*_{{\bf k},\uparrow\downarrow} + B^{(i)} u^*_{\bf k}t_{-{\bf k},\uparrow\uparrow}]/E_{\bf k} \\ +[A^{(i)} v^*_{\bf k}t^*_{{\bf k},\uparrow\uparrow} - B^{(i)} u^*_{\bf k}t_{-{\bf k},\uparrow\downarrow}]/E_{\bf k}\end{array}\right],
\end{equation}
($i=1,2$) where $A^{(i)}, B^{(i)}$ are $O(1)$. In this regime, we have
\begin{equation}
\langle f_{-{\bf k}\uparrow} f_{{\bf k}\uparrow} \rangle \approx [B^{(1)}]^* B^{(2)} = [A^{(1)} B^{(1)}]^*.
\end{equation}
Note that up to quadratic order in $t_{{\bf k}\sigma}$,  $(A^{(i)}, B^{(i)})^T$ are effectively eigenvectors of
\begin{equation}
H_{eff} = \frac{1}{E_{\bf k}}(-\tau^z \zeta_{\bf k} - \tau^x {\rm Re}\eta_{\bf k} + \tau^y {\rm Im} \eta_{\bf k}),
\end{equation}
when $|t_{{\bf k},\uparrow\sigma}|^2 = |t_{-{\bf k},\uparrow\sigma}|^2$. To justify this approximation we need
\begin{equation}
\frac{|t_{{\bf k},\uparrow\downarrow}|}{|t_{{\bf k},\uparrow\downarrow}|}\frac{ |u_{\bf k} v_{\bf k}|}{||u_{\bf k}|^2-|v_{\bf k}|^2|} \gg \frac{|t_{{\bf k},\uparrow\downarrow}|}{|\Delta_{\bf k}|}
\end{equation}
at the HM Fermi surface. Within this approximation we obtain
\begin{equation}
\langle f_{-{\bf k}\uparrow} f_{{\bf k}\uparrow} \rangle \approx \frac{1}{2}\frac{\eta_{\bf k}}{\sqrt{|\zeta_{\bf k}|^2 +|\eta_{\bf k}|^2}}.
\end{equation}
We also note that the half-metal spectrum is now gapped, with the gap of $\frac{\sqrt{|\eta_{\bf k}|^2 +|\zeta_{\bf k}|^2}}{E_{\bf k}}$. However, it is clear that only the $\eta_{\bf k}$ part is related to pairing, so we obtain
\begin{equation}
\Delta^{HM}_{\bf k} = \frac{\eta_{\bf k}}{E_{\bf k}}.
\end{equation}

\subsection{Strong hopping at the HM Fermi surface}

In order to apply perturbation theory in this limit, we note that $\Delta_{\bf k}/t_{{\bf k},\uparrow\uparrow}$ and $t_{{\bf k},\uparrow\downarrow}/t_{{\bf k},\uparrow\uparrow}$ are small variables now. 
It is therefore convenient to rewrite $H^{BdG}_{\bf k}$ in a new basis consisting of a spin down electron and a spin down hole in the $s$-wave SC and spin up electrons and holes in bonding and anti-bonding states over the HM and the $s$-wave SC. At the HM Fermi surface, this gives us
\begin{equation}
\tilde{H}^{BdG}_{\bf k} = U^\dagger H^{BdG}_{\bf k} U \nonumber\\
                                  = \left[\begin{array}{cc}
                                                      \tilde{h}^{tun}_{\bf k}  & (\tilde{T}^{pair}_{\bf k})^\dagger\\
                                                      \tilde{T}^{pair}_{\bf k} & \tilde{\Lambda}_{\bf k}
                                       \end{array}\right],
\end{equation}
where $\tilde{\Lambda}_{\bf k} = \tau^z (|u_{\bf k}|^2 - |v_{\bf k}|^2)E_{\bf k}$,
\begin{align}
\tilde{h}^{tun}_{\bf k} &= {\rm diag}(t_{{\bf k},\uparrow\uparrow}, t_{-{\bf k},\uparrow\uparrow}, -t_{{\bf k},\uparrow\uparrow}, -t_{-{\bf k},\uparrow\uparrow})\nonumber\\
&+ \frac{|u_{\bf k}|^2 - |v_{\bf k}|^2}{2}E_{\bf k}\left[\begin{array}{cccc}
1 &  & -1 & \\
 &  -1 &  & 1\\
-1&  & 1  & \\
 & 1 &  &  -1\end{array}\right],
 \end{align}
 \begin{equation}
\tilde{T}^{pair}_{\bf k} = \sqrt{2}\left[\begin{array}{cccc}
t^*_{{\bf k},\uparrow\downarrow}/2 & u^*_{\bf k} v_{\bf k} E_{\bf k} & t^*_{{\bf k},\uparrow\downarrow}/2 & -u^*_{\bf k} v_{\bf k} E_{\bf k}\\
u_{\bf k} v^*_{\bf k} E_{\bf k} & -t_{-{\bf k},\uparrow\downarrow}/2   & -u_{\bf k} v^*_{\bf k} E_{\bf k} & -t_{-{\bf k},\uparrow\downarrow}/2
\end{array}\right],
\end{equation}
and
\begin{equation}
U = \left[\begin{array}{cccccc}
 \frac{1}{\sqrt{2}} &  &  \frac{1}{\sqrt{2}}& & & \\
  & \frac{1}{\sqrt{2}} &  & \frac{1}{\sqrt{2}} & & \\
\frac{u_{\bf k}}{\sqrt{2}} & &-\frac{u_{\bf k}}{\sqrt{2}} & & & -v_{\bf k}\\
 & -\frac{v_{\bf k}}{\sqrt{2}} &  & \frac{v_{\bf k}}{\sqrt{2}} & u_{\bf k} &\\
 & -\frac{u^*_{\bf k}}{\sqrt{2}} &  & \frac{u^*_{\bf k}}{\sqrt{2}}  & -v^*_{\bf k} &\\
\frac{v^*_{\bf k}}{\sqrt{2}} & &-\frac{v^*_{\bf k}}{\sqrt{2}} & & & u^*_{\bf k}
\end{array}\right]
\end{equation}
(note that chose a gauge so that $t_{{\bf k},\uparrow\uparrow}$ is real positive).

Our strong-hopping mode in 2Dl gives us three spin-polarized bands. We can see this from examining the column vectors of $U$. Note that the first and third columns represents the electron in the spin-up bonding and anti-bonding bands respectively, the second and fourth columns represents the hole in the spin-up bonding and anti-bonding bands respectively, and the last two columns represents the electron and the hole in the spin-down band. For $t_{{\bf k},\uparrow\uparrow} = t_{-{\bf k},\uparrow\uparrow}$, we have three pairs of a electron state and a hole state degenerate in energy - two pairs for spin-up and one pair for spin-down. Therefore, we can again apply degenerate second order perturbation theory with three sets of bases
\begin{equation}
{\bf \tilde{F}}^{(i)}_{\bf k} = \left[\begin{array}{c}
A^{(i)}\\
B^{(i)}\\
-A^{(i)} \frac{|u_{\bf k}|^2 - |v_{\bf k}|^2}{4}\frac{E_{\bf k}}{t_{{\bf k},\uparrow\uparrow}}\\
B^{(i)} \frac{|u_{\bf k}|^2 - |v_{\bf k}|^2}{4}\frac{E_{\bf k}}{t_{-{\bf k},\uparrow\uparrow}}\\
A^{(i)} \frac{1}{\sqrt{2}}\frac{t^*_{{\bf k},\uparrow\downarrow}}{t_{{\bf k},\uparrow\uparrow}} + B^{(i)} \sqrt{2}u^*_{\bf k} v_{\bf k} \frac{E_{\bf k}}{t_{-{\bf k},\uparrow\uparrow}}\\
A^{(i)} \sqrt{2}u_{\bf k} v^*_{\bf k} \frac{E_{\bf k}}{t_{{\bf k},\uparrow\uparrow}} - B^{(i)}\frac{1}{\sqrt{2}}\frac{t_{-{\bf k},\uparrow\downarrow}}{t_{-{\bf k},\uparrow\uparrow}}
\end{array}\right],
\end{equation}
\begin{equation}
{\bf \tilde{G}}^{(i)}_{\bf k} = \left[\begin{array}{c}
C^{(i)}\frac{|u_{\bf k}|^2 - |v_{\bf k}|^2}{4}\frac{E_{\bf k}}{t_{{\bf k},\uparrow\uparrow}}\\
-D^{(i)}\frac{|u_{\bf k}|^2 - |v_{\bf k}|^2}{4}\frac{E_{\bf k}}{t_{-{\bf k},\uparrow\uparrow}}\\
C^{(i)}\\
D^{(i)}\\
-C^{(i)} \frac{1}{\sqrt{2}}\frac{t^*_{{\bf k},\uparrow\downarrow}}{t_{{\bf k},\uparrow\uparrow}}+D^{(i)} \sqrt{2}u^*_{\bf k} v_{\bf k} \frac{E_{\bf k}}{t_{-{\bf k},\uparrow\uparrow}}\\
C^{(i)} \sqrt{2}u_{\bf k} v^*_{\bf k} \frac{E_{\bf k}}{t_{{\bf k},\uparrow\uparrow}}+D^{(i)}\frac{1}{\sqrt{2}}\frac{t_{-{\bf k},\uparrow\downarrow}}{t_{-{\bf k},\uparrow\uparrow}}
\end{array}\right],
\end{equation}
\begin{equation}
{\bf \tilde{K}}^{(i)}_{\bf k} = \left[\begin{array}{c}
-X^{(i)}\frac{1}{\sqrt{2}}\frac{t_{{\bf k},\uparrow\downarrow}}{t_{{\bf k},\uparrow\uparrow}}-Y^{(i)}\sqrt{2}u^*_{\bf k} v_{\bf k} \frac{E_{\bf k}}{t_{{\bf k},\uparrow\uparrow}}\\
-X^{(i)}\sqrt{2}u_{\bf k} v^*_{\bf k} \frac{E_{\bf k}}{t_{-{\bf k},\uparrow\uparrow}}+Y^{(i)}\frac{1}{\sqrt{2}}\frac{t^*_{-{\bf k},\uparrow\downarrow}}{t_{-{\bf k},\uparrow\uparrow}}\\
X^{(i)}\frac{1}{\sqrt{2}}\frac{t_{{\bf k},\uparrow\downarrow}}{t_{{\bf k},\uparrow\uparrow}}-Y^{(i)}\sqrt{2}u^*_{\bf k} v_{\bf k} \frac{E_{\bf k}}{t_{{\bf k},\uparrow\uparrow}}\\
-X^{(i)}\sqrt{2}u_{\bf k} v^*_{\bf k} \frac{E_{\bf k}}{t_{-{\bf k},\uparrow\uparrow}}-Y^{(i)}\frac{1}{\sqrt{2}}\frac{t^*_{-{\bf k},\uparrow\downarrow}}{t_{-{\bf k},\uparrow\uparrow}}\\
X^{(i)}\\
Y^{(i)}
\end{array}\right].
\end{equation}

We can show from the effective Hamiltonians for the above three bases that the chirality of the spin-down pairs will be opposite to that of the two sets of spin-up pairs. The effective Hamiltonians of ${\bf \tilde{F}}^{(i)}_{\bf k}$ and ${\bf \tilde{G}}^{(i)}_{\bf k}$ can be written
\begin{align}
\tilde{H}^\pm_{eff,\uparrow} \approx& \frac{E_{\bf k}}{|t_{{\bf k},\uparrow\uparrow}|^2}(\tau^z \zeta_{\bf k} + \tau^x {\rm Re}\eta_{\bf k} - \tau^y {\rm Im} \eta_{\bf k})\nonumber\\
&\pm |t_{{\bf k},\uparrow\uparrow}|\left[\frac{1}{2}\frac{|\Delta_{\bf k}|^2}{|t_{{\bf k},\uparrow\uparrow}|^2}+ \frac{1}{2}\frac{|t_{{\bf k},\uparrow\downarrow}|^2}{|t_{{\bf k},\uparrow\uparrow}|^2}+\frac{1}{8}\frac{(\epsilon'_{\bf k} - \mu')^2}{|t_{{\bf k},\uparrow\uparrow}|^2}\right],
\label{EQ:effectiveUp}
\end{align}
while for ${\bf \tilde{K}}^{(i)}_{\bf k}$, the effective Hamiltonian is
\begin{equation}
\tilde{H}_{eff,\downarrow} \approx \frac{2E_{\bf k}}{|t_{{\bf k},\uparrow\uparrow}|^2}(\tau^z \zeta_{\bf k} - \tau^x {\rm Re}\eta_{\bf k} -\tau^y {\rm Im} \eta_{\bf k}).
\label{EQ:effectiveDown}
\end{equation}
Because the tunneling process required for pairing the spin-down pairs is the reverse of what is required for pairing the spin-up pairs, the chirality of the spin-down pairs is opposite of that of the spin-up pairs. Eqs.\eqref{EQ:effectiveUp} and \eqref{EQ:effectiveDown} show that the HM / SC heterostructure overall has the $\mathcal{N} = 1$ TSC. We believe that this result should hold even in the case the SC is 3D.

These effective Hamiltonian tells us that we have giving us
\begin{equation}
\Delta^{HM}_{\bf k} \approx \frac{\eta_{\bf k}}{|t_{{\bf k},\uparrow\uparrow}|^2}\Delta_{\bf k},
\end{equation}
since only the spin-up pairs involve the HM.

\section{Polar catastrophe in zinc-blende ${\rm {\bf VTe/ZnTe}}$ (111)}

In this Appendix, we discuss the physical origin of the ``polar catastrophe" in the zinc-blende VTe/ZnTe (111) that gives us a single Fermi surface at 0.0eV. To explain this, we first need to clarify what crystal structure we used for the band structure calculation. 
To maintain the zinc blende structure of VTe, we always need a zinc blende substrate with a good matching lattice constant. The zinc blende ZnTe not only has the lattice constant matching within 1$\%$ but also has been used experimentally in growing the zinc blende CrTe. For VTe, growing either of them on the (111) side of the zinc blende ZnTe is sufficient in obtaining both half-metallicity with a single Fermi surface and the two atomic layer zinc blende (111) structure with very little structural distortion. The crystal structure of this heterostructure can be a series of monatomic triangular lattice in the sequence V-Te-Zn-Te-Zn-Te-... with all Zn atoms sitting at the center of tetrahedron formed by Te atoms while Te atoms adjacent to V is one short of forming tetrahedron. (The zinc-blende CrTe/ZnTe (111) has essentially the same crystal structure.)

We now explain how we effectively have three and one-half electrons per unit cell of the VTe at the top. This is crucial because having a fractional number of electron per unit cell is necessary for a half-metal to have a single Fermi pocket at the Fermi level. We first note that ZnTe is an insulator (with a gap of 2.3 eV) due to covalent bonding between Zn and Te takes up all valence electrons of this atom, with Zn contributing 2 electrons per atom, or equivalently, 2/4 = 1/2 electrons per bonding with Te. We need to note here that the electron configuration of a V atom differ from that of a Zn atom due to addition of three $3d$ electrons. Since each V atom is missing one Te atom, we get 1/2 electrons for every dangling bond from V. While these electrons from dangling bonds cannot flow into the bulk ZnTe, the VTe film is metallic due to V $3d$ electrons. In short, the dangling bonds and $3d$ electrons conspire to give us a half-metal with a single Fermi pockets, and it is from these four orbitals that we get one partially filled bands and three filled bands close to the Fermi level in Fig 4 (a). 
As we have pointed out, this is analogous to the ``polar catastrophe" at the LaAlO$_3$ / SrTiO$_3$ interface.

We have set up our calculation so that this ``polar catastrophe" would not lead to any breakdown. In our calculation, when we used a ZnTe of finite thickness, the dipole moment between Zn and Te atoms give rises to an electric field in our heterostructure. Since this effect decreases when we increased the thickness of the ZnTe layer, we expect the effect to go away for a sufficiently thick ZnTe layer. For our band calculation, we added an extra atomic layer of the H atoms - which gives a VTe/ZnTe/H heterostructure - solely to cancel out this artificial electric field. In Fig 4 (a) we have omitted the H band, which do not mix with any VTe band due to both the insulating gap and the thickness of the ZnTe layer.

\medskip

\section{The method for {\it ab-initio} calculations}

In this work, all the {\it ab-initio} calculations are performed by the BSTATE(Beijing Simulation Tool of
Atomic Technology) package with the plane-wave pseudopotential method. The Perdew-Burke-Ernzerhof type\cite{perdew1996} generalized gradient approximation(GGA) is used for exchange-correlation energy. Due to 2D models, the $\mathbf{k}$-point grid is taken as 12$\times$12$\times$1 for general self-consist calculations, and the kinetic energy cutoff is fixed to 340eV. All the atoms on the surface part are fully relaxed for our free-standing slab zb-VTe/ZnTe (111) model. However, for our CrO$_2$ model, we assumed that the perfect rutile structure and applied no relaxation.


\end{document}